\documentclass[sigconf]{acmart}

\usepackage{chapterbib}
\usepackage{docmute}

\usepackage{graphicx}
\usepackage{subcaption}
\usepackage{multirow}

\usepackage{stfloats}
\usepackage{makecell}
\usepackage{threeparttable}

\usepackage{algorithm}
\usepackage{algpseudocode}
\usepackage{hyperref}
\hypersetup{hidelinks,
	colorlinks=true,
	allcolors=blue,
    urlcolor=cyan,
	breaklinks=true}

\AtBeginDocument{%
  }

\copyrightyear{2025}
\acmYear{2025}
\setcopyright{cc}
\setcctype{by}
\acmConference[FSE Companion '25]{33rd ACM International Conference on the Foundations of Software Engineering}{June 23--28, 2025}{Trondheim, Norway} \acmBooktitle{33rd ACM International Conference on the Foundations of Software Engineering (FSE Companion '25), June 23--28, 2025, Trondheim, Norway}
\acmDOI{10.1145/3696630.3728546}
\acmISBN{979-8-4007-1276-0/2025/06}

\begin{document}

\title{Grey-Box Fuzzing in Constrained Ultra-Large Systems: Lessons for SE Community}

\author{Jiazhao Yu}
\affiliation{
  \institution{Sun Yat-sen University}
  \city{Guangzhou}
  \country{China}
}
\email{yujzh7@mail2.sysu.edu.cn}

\author{Yanlun Tu}
\affiliation{
  \institution{Ant Group}
  \city{Shanghai}
  \country{China}
}
\email{tuyanlun.tyl@antgroup.com}

\author{Zhanlei Zhang}
 \affiliation{
  \institution{Macquaire University}
  \city{Sydney}
  \country{Australia}
}
\email{zzlbirenoulu@gmail.com}

\author{Tiehua Zhang}
\authornotemark[1]
\affiliation{
  \institution{Tongji University}
  \city{Shanghai}
  \country{China}
}
\email{tiehuaz@tongji.edu.cn}

\author{Cheng Xu}
\affiliation{
  \institution{Ant Group}
  \city{Shanghai}
  \country{China}
}
\email{kunqiu.xc@antgroup.com}

\author{Weigang Wu}
\authornotemark[1]
\affiliation{
  \institution{Sun Yat-sen University}
  \city{Guangzhou}
  \country{China}
}
\email{wuweig@mail.sysu.edu.cn}

\author{Hong Jin Kang}
\affiliation{
  \institution{University of Sydney}
  \city{Sydney}
  \country{Australia}
}
\email{hongjin.kang@sydney.edu.au}

\author{Xi Zheng}
\authornote{Corresponding authors: Tiehua Zhang, Weigang Wu and Xi Zheng.}
\affiliation{
  \institution{Macquaire University}
  \city{Sydney}
  \country{Australia}
}
\email{james.zheng@mq.edu.au}

\renewcommand{\shortauthors}{Yu et al.}

\begin{abstract}
Testing ultra-large microservices-based FinTech systems presents significant challenges, including restricted access to production environments, complex dependencies, and stringent security constraints. We propose \textsc{SandBoxFuzz}, a scalable grey-box fuzzing technique that addresses these limitations by leveraging aspect-oriented programming and runtime reflection to enable dynamic specification mining, generating targeted inputs for constrained environments. \textsc{SandBoxFuzz} also introduces a log-based coverage mechanism, seamlessly integrated into the build pipeline, eliminating the need for runtime coverage agents that are often infeasible in industrial settings. 
\textsc{SandBoxFuzz} has been successfully deployed to Ant Group's production line and, compared to an initial solution built on a state-of-the-art fuzzing framework, it demonstrates superior performance in their microservices software.
\textsc{SandBoxFuzz} achieves a 7.5\% increase in branch coverage, identifies 1,850 additional exceptions, and reduces setup time from hours to minutes, highlighting its effectiveness and practical utility in a real-world industrial environment.
By open-sourcing \textsc{SandBoxFuzz}\footnote{\href{https://zenodo.org/records/14679560}{https://zenodo.org/records/14679560}}, we provide a practical and effective tool for researchers and practitioners to test large-scale microservices systems.
\end{abstract}





\maketitle

\section{Introduction}

The growing complexity of modern software systems has significantly increased the prevalence of software vulnerabilities, which can compromise operational efficiency and even lead to catastrophic failures, particularly in large-scale systems \cite{DBLP:journals/pieee/LinWHZX20}. Fuzz testing, or fuzzing, has emerged as a powerful automated technique for identifying erroneous behaviors in diverse software programs \cite{DBLP:journals/cacm/MillerFS90, DBLP:conf/ccs/BohmePR16, DBLP:journals/csur/ZhuWCX22, yun2022fuzzing, zhang2021intelligen, lipner2004trustworthy}. Existing fuzzing tools have proven highly effective at detecting software defects in real-world applications \cite{americanfuzzylop, AFLPP, libfuzzer, zest}. By identifying and mitigating security vulnerabilities, fuzzing plays a crucial role in reducing their impact and minimizing long-term financial costs. Fuzzing has also gained significant traction within industry. Industrial studies \cite{aizatsky2016announcing, liang2018fuzz} have developed powerful fuzzing tools capable of uncovering bugs in various programs (e.g., internal middleware) or accelerating the fuzzing process through parallel processing. However, the feasibility of applying fuzz testing within constrained industrial environments remains largely unexplored, especially where fuzzers can be difficult to integrate and scale, due to organization-wide security constraints, and the opacity and scale of the software systems. These unique issues have yet to be thoroughly discussed or addressed in existing research. 

Through our practical research and development experience in Ant Group, a prominent FinTech company, we observe three main challenges in deploying fuzz testing to large-scale FinTech systems:

\begin{itemize}
\item \textbf{Limited Permissions.} 
Due to data security concerns, access to repositories and source code are restricted. Users of the fuzzers are not necessarily the authors of code in a target repository. They may not have full access to the repositories as well. Therefore, white-box-based fuzzing approaches \cite{zhang2023EvoMaster,10.1145/3652157,gupta2024robust,strassle2024systematic,10771466,LIU2025107640,10.1145/3689736,arcuri2025tool,YourFixIsMyExploit} that require access to source code are infeasible. Enabling the fuzzer users to rapidly set up a runnable fuzzing pipeline is a key factor for scalable deployment.
\item \textbf{Restricted Test Environments.} While performing testing within an enterprise's settings, unlike operating in a fully controlled environment, test cases are enforced to execute using designated test frameworks and toolsets. These tools facilitate testing processes for developers. 
Due to their black-box nature, modifications to both the test environment and framework are strictly prohibited. While fuzzers rely on a tight feedback loop, the test environment restricts access to critical feedback, such as coverage information obtained through runtime agents. In Ant Group, the only permissible action is fetching logs from the test framework, which can only be done after the execution process of the entire test suite has been terminated.
The existing corporate testing pipelines requires restarting the application for each fuzzing iteration, resulting in significant cold start overhead. This increases the challenges of integrating fuzz testing into industrial software systems.
\item \textbf{Constraints from Input Constructions.} Test inputs have various formats and complex compositions, which are introduced by designated test frameworks that are enforced by the company’s policies. File-based formats are ease-of-use and the most popular ones among developers. However, when combining with unique features of the test framework, additional overhead and constraints are introduced, including large-scale discard of inputs, limited mutation space, overhead of file operations and expensive manual effort on constructing mutation specifications. 
\end{itemize}

To tackle these challenges, in this paper, we propose \textsc{SandBoxFuzz}, a coverage-guided grey-box fuzzer for Java that bridges the gaps between generic fuzzing methodologies and practical deployment in industrial settings. Our improvements in \textsc{SandBoxFuzz} are based on three insights: First, the long startup cost of the application can be amortized by launching many test cases in one iteration. Second, bytecode can be augmented since it complies with data security regulations and produces useful metrics for guiding grey-box fuzzing, without breaking the black-box test environment. Third, substantial preparation time can be saved by replacing manual specification construction with automatic specification mining.

As a result, rather than configure test input data and start the slow testing workflow, we utilize a sandbox approach to intercept inputs at specific \textit{pointcuts} during the actual testing process, allowing us to bypass many of the restrictions imposed by existing test configurations. Specifically, we bootstrap the fuzzing process by launching the testing pipeline with duplicated valid inputs, while these dummy inputs will be replaced with real ones at runtime. Furthermore, bytecode of target program can be instrumented with print statements to track execution of each basic block in control-flow graph (CFG). Finally, we can save huge amount of manual effort, from hours to minutes, by mining specifications from the execution of the manually written test cases. Combining the observations and improvements, \textsc{SandBoxFuzz} makes fuzz testing deployable and scalable under the industrial context of Ant Group, with minimal manual effort. 

We apply \textsc{SandBoxFuzz} in Ant Group by integrating the fuzzing pipeline into its test environment. We evaluate \textsc{SandBoxFuzz} on target \textit{entry method}s selected from several applications that are part of an ultra-large scale microservices system. Our experimental results show that \textsc{SandBoxFuzz} outperforms the manual testing approach at fuzzing effectiveness, and consumes less overall time including preparation and execution processes, with acceptable overhead introduced by its components. Moreover, \textsc{SandBoxFuzz} is better suited for an industrial environment where the testing framework is fixed and not optimized for fuzzing.
It also stands out as a tool for users with limited knowledge of fuzzing to perform agile fuzz testing as it is the practical solution in terms of minimizing manual effort and scaling the fuzzing pipeline.

In summary, this paper makes the following contributions:
\begin{itemize}
\item We highlight the unique challenges and constraints arising from the application of fuzz testing in real industrial systems with restricted setups, based on our real practice in Ant Group. To the best of our knowledge, we are the first to raise these issues, bridging the gap between deploying efficient and scalable fuzzers, and operating within constrained conditions.
\item We have developed \textsc{SandBoxFuzz}, a coverage-guided grey-box fuzzer suited for common industrial test settings. \textsc{SandBoxFuzz} has solved or bypassed various restrictions imposed by these settings. It also facilitates a scalable fuzzing deployment in real industrial context by saving substantial manual effort in setting up the fuzzing pipeline.
\item We have deployed \textsc{SandBoxFuzz} on 4 applications of real-world microservices software and evaluated them on a total of 37 \textit{entry methods}. Compared with the baseline approach, \textsc{SandBoxFuzz} improves by 6.0\% and 7.5\% in method and branch coverage, respectively. \textsc{SandBoxFuzz} yields 1,850 more cases that trigger the target exceptions. 
In our assessment of its scalability, \textsc{SandBoxFuzz} only requires an average of 9 minutes for its set up to target a new \textit{entry method}. In contrast, configuring the file-based approach requires several hours. This demonstrates that \textsc{SandBoxFuzz} can be deployed practically 
in production.
\end{itemize}

\section{BACKGROUND} \label{background}

In this section, we first explore the common characteristics of testing in an enterprise, taking Ant Group as an example. 
We present challenges and constraints encountered in our real practice. 
These issues, which are unique and representative, have received limited attention yet can significantly impact the design, implementation, and deployment of fuzz testing within industrial settings.

\begin{figure}[h]
    \centering
    \includegraphics[width=1\linewidth]{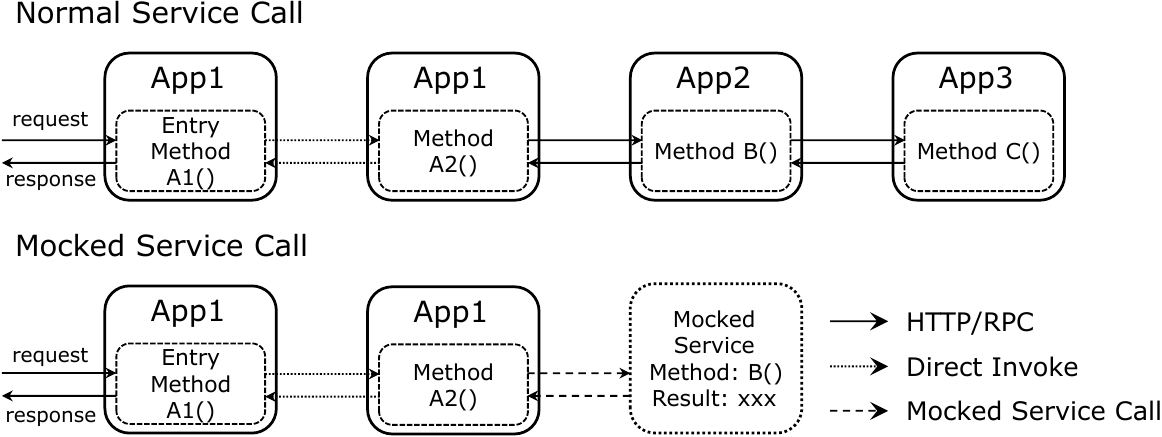}
    \caption{\textbf{An example of microservices service calls.}}
    \Description{}
    \label{fig:service}
\end{figure}

\subsection{Knowledge of Testing in Enterprise} \label{sub:tie}
\textbf{System Under Test.}
Software systems in enterprises typically adopt a microservices architecture, where services are divided into a collection of independent components (or applications) that communicate with one another through network protocols such as HTTP and RPC. 
Specifically, business logics involve service calls between multiple applications, as shown in Figure \ref{fig:service}. 
To test the functionality of each application that depends on other components, developers have to either construct mocks for the components, which return pre-defined results, or rely on the deployment of the components into a staging or testing environment. The loosely coupled nature of microservices allows each team to independently develop, deploy, and maintain respective application components. 
However, this gives rise to testing issues related to non-determinism, i.e., different results may be returned upon invocations at different times, due to continuous software updates and deployments into the staging environment. Most of the components are developed in Java and leverage the widely-used Spring framework \cite{spring}. They are integrated into a continuous integration and deployment (CI/CD) pipeline and are deployed and launched via SOFABoot \cite{sofaboot}, an enhanced implementation of SpringBoot that provides features such as readiness check and context isolation, which are helpful for microservices deployment.

\textbf{Scale of the Software.}
Currently, Ant Group owns multiple microservices systems. Each system comprises from hundreds to thousands of application components. Each application is complex, with sizes exceeding thousands of megabytes and over 1 million LoC (lines of code). Each application is accompanied by a test suite written manually by the developers. 
A manual test suite usually contains hundreds to thousands of hand-written test cases, linked to dozens to hundreds of unique \textit{entry methods}.

\begin{figure}[htbp]
    \centering
    \Description{}
    \begin{subfigure}[t]{0.38\textwidth}
        \centering
        \includegraphics[width=\textwidth]{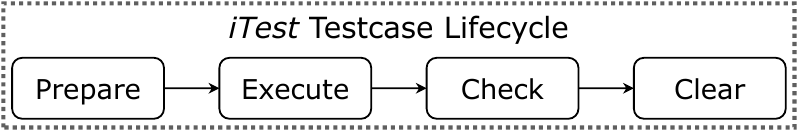}
        \caption{\textbf{Lifecycle of a single test logic.}}
        \label{fig:itest-lifecycle}
    \end{subfigure}
    \vfill
    \begin{subfigure}[t]{0.38\textwidth}
        \centering
        \includegraphics[width=\textwidth]{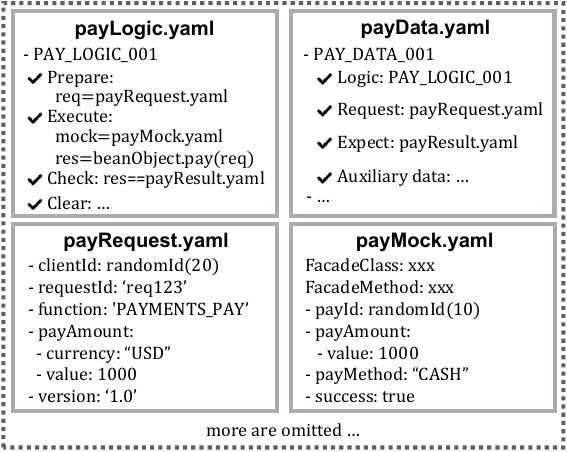}
        \caption{\textbf{Complex composition of test cases.}}
        \label{fig:itest-case}
    \end{subfigure}
    \caption{\textbf{Main features of \textit{iTest} test framework.}}
    \label{fig:itest}
\end{figure}

\textbf{Test Environments.}
Enterprise-level systems typically have designated test frameworks and toolsets, including well-known ones such as Junit and TestNG or custom ones. In this paper, we develop and deploy a fuzzer on a test framework widely adopted in Ant Group, namely \textit{iTest}, which is built on TestNG. \textit{iTest} works within the test lifecycle of the project management tool and integrates most of the test types (e.g., unit testing, component testing, and integration testing). \textit{iTest} offers greater ease-of-use to developers by allowing developers to focus on configuring the test inputs rather than configuring the test environment. Since \textit{iTest} performs the setup of the complex test environment with a large number of dependencies, e.g., service discovery and module loading, it would be challenging for developers to set up a valid test environment without the use of \textit{iTest}. 
However, as described in Section \ref{sub:challenges}, the design of \textit{iTest} introduces challenges for the integration and deployment of a fuzzer. 

To increase modularity, \textit{iTest} has decoupled the test data and test logic. Each unique pair of them forms a unique test case.
As shown in Figure \ref{fig:service}, two sets of configuration form the input data for a single test case: the request objects, and upstream or downstream service calls (which can be replaced by mocks). Statements in test logic defines the target \textit{entry method} and behaviors throughout its lifecycle. Figure \ref{fig:itest-lifecycle} illustrates the main lifecycle of a single test logic in \textit{iTest}, which comprises four primary stages as follows:

\begin{itemize}
\item \textbf{Preparation.} This phase establishes a specific test context for the current test case, constructs any necessary request objects from files and prepares for any additional data.
\item \textbf{Execution.} This phase first sets up the mock components from files (if any), then retrieves the target bean object from the Spring container, and finally invokes the target \textit{entry method} on the object with given request data.
\item \textbf{Check.} This phase compares the actual results against the expected ones (if declared), which are defined in the manual test suite.
\item \textbf{Clear.} This phase denotes the end of the test logic and removes any residual data from the test context.
\end{itemize}

File-based formats (e.g., \texttt{yaml} and \texttt{json}) are used in component or integration testing. 
In \textit{iTest}, a file-based test case consists of multiple parts.
As shown in Figure \ref{fig:itest-case}, the \texttt{payLogic.yaml} file defines the components for the \textit{iTest} lifecycle, including the mock files (\texttt{payMock.yaml}) and input parameters (\texttt{payRequest.yaml}), while auxiliary data is specified in \texttt{payData.yaml}.
These configuration files are parsed and converted into objects held in memory once \textit{iTest} starts testing and loads the test cases.

Overall, upon starting a testing process, \textit{iTest} boots up the application, loads the dependent services and components, collects test cases from files to build a test set, constructs any necessary data for the target test set, and executes through the same test lifecycle for each test case.

\subsection{Challenges} \label{sub:challenges}
There are some common assumptions when implementing a generic fuzzer. However, they do not hold in the settings of a large-scale software system. We have identified several critical gaps in our practice between theoretical assumptions and real-world conditions.

\textbf{Assumption 1:} Operations are granted with full access to source code and full control over test environments. 

\textbf{Reality \& Challenge 1:} FinTech companies have strict data security regulations. Due to these concerns, global source code access is considered more sensitive than access to bytecode.
Thus, only restricted access is granted to a small subset of codebase resources, making the white-box approaches infeasible. Limited permissions and inability to analyze the source code of the target application also lead to a lack of sufficient knowledge about target software. Meanwhile, due to microservices settings, the use of test frameworks is enforced (here, \textit{iTest}). The black-box nature of the test framework prohibits any access to a fuzzer's executor and monitor, which makes it difficult to provide a fuzzer with feedback apart from logs from print statements generated by
the test framework after execution has terminated.

\textbf{Assumption 2:} Fuzz testing should be fast and efficient, that is, both execution and iteration should be fast. 

\textbf{Reality \& Challenge 2:} Unfortunately, in some industrial settings, the execution of even a single test case may not be fast.  
For input execution, the test framework processes through an independent and entire lifecycle for every single test case. In \textit{iTest}, completing the lifecycle can take 
up to a few seconds, depending on multiple factors such as the complexity of inputs and test logics, and current service load in the test environment. For each fuzzing iteration, the test framework has to launch the application and load the service components before processing the inputs from test set. For \textit{iTest}, this process can take more than several minutes, depending on the scale and number of the modules included in the target application. 
Furthermore, the lifecycle of testing pipeline does not last indefinitely and the pipeline terminates once the test framework has executed all existing inputs. Thus, given updates on input files, the test framework has to reboot the application to initiate a new test lifecycle and load the new inputs from files. Hot-swapping technique ~\cite{hotswap}, which can update Java classes in place and prevent the need to reload the container, is only supported for reloading bytecode but not for the input files.
Hence, unlike a generic fuzzing loop where inputs can be generated and loaded dynamically, each fuzzing iteration must go through a cold start process within the industrial settings. Compared to generic fuzzers, where at least hundreds or thousands of inputs are generated and executed per second, this issue significantly limits the speed and efficacy of an industrial fuzzing pipeline.

\textbf{Assumption 3:} Configuring test inputs as file-based inputs provides ease-of-use without additional overhead or side-effects. 

\textbf{Reality \& Challenge 3:} In industrial testing, developers prefer configuring test data in file (\texttt{yaml} and \texttt{json} files) due to its ease-of-use and greater modularity. \textit{iTest} provides file-based configurations and is popular within the organization because it removes the responsibility of setting up the test environment from users. 
\textit{iTest} transforms the inputs configured in the \texttt{yaml} and \texttt{json} files into Java objects held in memory. However, when encountering failures during transforming the inputs (e.g., \textit{syntactically invalid}), \textit{iTest} abandons the execution of the whole test set, raising an error to shorten the feedback loop for developers as they can quickly respond to the failure and fix the incorrect input configurations. 
These features make it convenient for developers to perform testing upon any new added logics. However, for implementation of a fuzzer, these features raise issues. First, any single \textit{syntactically invalid} input causes the entire test set to be discarded and wastes computational and manual effort on validity checking. 
Mutation space is also limited, because certain input values (e.g., escape characters) and complex objects are illegal or difficult to be expressed in file formats (taking too much space and not being efficient).
Next, \textit{iTest} searches through the input files for target values in multiple stages including input data parsing and result validation. With each value occupies a different size in files, the search has an unpredictable overhead on file operations. 
At last, input specifications are required for mutating inputs into other valid inputs, as each input has a different type and range of valid values. 
Note that the lack of specification and its construction process are crucial issues as they impose non-trivial manual effort to construct input specifications for each target \textit{entry method}, raising the difficulty of scaling the fuzzer to different entry methods and applications. 
Overall, while the features of \textit{iTest} ease the work of developers manually writing test cases, they introduce challenging constraints for the implementation of a scalable fuzzer that can be deployed within the industrial testing pipeline across multiple applications.

\section{APPROACH} \label{approach}

In this section, we introduce both \textsc{FatFuzz} and \textsc{SandBoxFuzz}.
\textsc{FatFuzz} is our initial solution deployed successfully at Ant Group. \textsc{SandBoxFuzz} is developed to address the limitations of \textsc{FatFuzz}. We describe the details of \textsc{SandBoxFuzz}, including architecture, workflow, and solutions to the challenges outlined in Section \ref{sub:challenges}, along with the differences between the two approaches.

\begin{figure}[htbp]
    \centering
    \includegraphics[width=0.9 \linewidth]{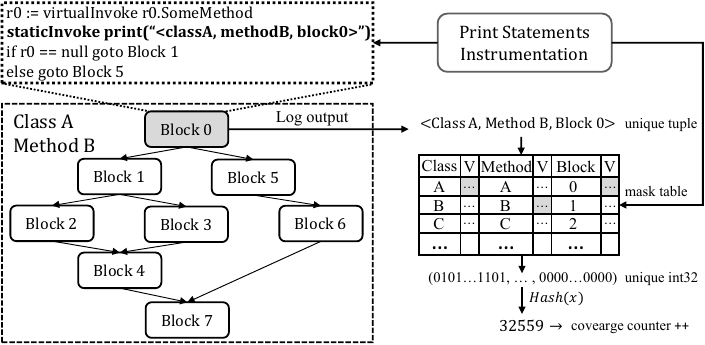}
    \caption{\textbf{An example of instrumenting print statements.}}
    \Description{}
    \label{fig:loginst}
\end{figure}

\subsection{The Predecessor --- \textsc{FatFuzz}}
In our early practice, we implemented a coverage-guided grey-box fuzzer within Ant Group's testing pipeline. 
Following state-of-the-art fuzzers \cite{zest,reddy2020quickly,nguyen2022bedivfuzz} and JQF \cite{JQF}, an open-source framework for coverage-guided testing in Java, we built our first solution, \textsc{FatFuzz}.

\begin{figure*}[htbp]
    \centering
    \includegraphics[width=1\linewidth]{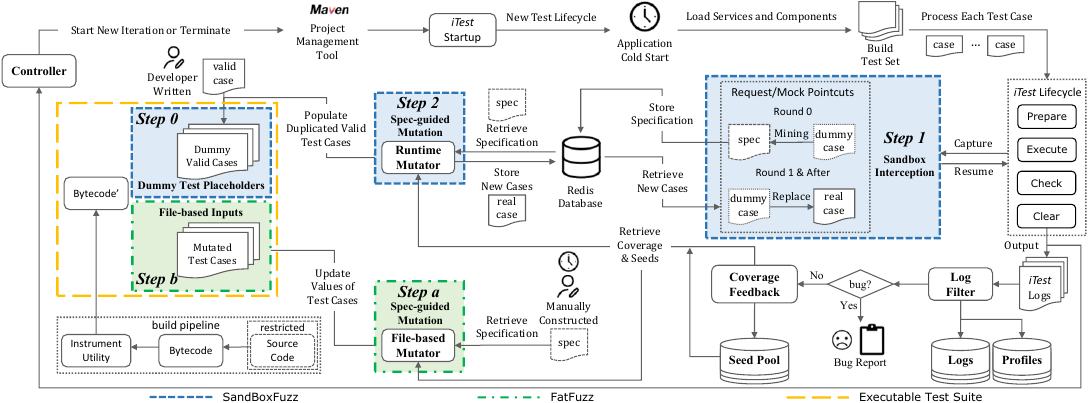}
    \caption{\textbf{Overview of \textsc{SandBoxFuzz}'s architecture and workflow.}}
    \Description{}
    \label{fig:SandboxFuzz}
\end{figure*}

\textit{iTest} provides the fuzzer with only limited information due to company-wide security policies, and its black-box nature prevents modifications that may leak information. 
Instrumentation approaches typically use a Java agent to augment the bytecode and collect coverage metrics at runtime for fuzzers. 
Meanwhile, as mentioned in Section \ref{sub:challenges}, at the start of the fuzzing process, a costly cold start is required in rebooting the application for loading the next set of test inputs.
However, this reset of the entire test environment also causes the loss of the coverage states. 
We make two designs to address these challenges, so as to make \textsc{FatFuzz} deployable and effective. 
First, \textsc{FatFuzz} executes a large number of inputs, rather than just a single one, in each invocation of the test framework, effectively amortizing the cost incurred by application cold starts.
Next, the application code is instrumented by an instrumentation utility integrated into the build pipeline as shown in Figure \ref{fig:SandboxFuzz}. 
This inserts print statements using the instrumentation tool provided by Zhao et al. \cite{log20}, so as to retrieve execution status from test environment and convert the printed statements present in the output of the actual test execution processes into coverage metrics. As shown in Figure \ref{fig:loginst}, using static analysis, control-flow graphs (CFGs) are constructed to identify independent code branches. For each class, method, and block, unique binary encodings are generated for all combinations, forming a mask table. Print statements with class, method, and block names are inserted into each basic block. When these statements appear in \textit{iTest} logs, the mask table is used to encode them, and a hash function generates a low-collision hash value. A coverage counter for each class-method-block pair is incremented by 1 each time the hash value is generated.

However, while \textsc{FatFuzz} solves the first two challenges in Section \ref{sub:challenges}, it does not address the constraints of Challenge 3. 
Critically, \textsc{FatFuzz} incurs significant manual effort, requiring input specifications for each target \textit{entry method} to avoid generating \textit{syntactically invalid} inputs. These specifications define input structures, field types, and valid ranges, but configuring them is error-prone, averaging several hours. Additionally, locating and formatting input values demands substantial effort. Despite its success in improving code coverage, \textsc{FatFuzz} is neither scalable nor practical, as developers are reluctant to adopt a fuzzer with extensive preparation requirements.
Meanwhile, \textsc{FatFuzz} has a limited mutation space, as keeping certain input values and complex objects are illegal or less efficient in files. These issues motivate the design of \textsc{SandBoxFuzz}.

\begin{figure}[htbp]
    \centering
    \includegraphics[width=0.57\linewidth]{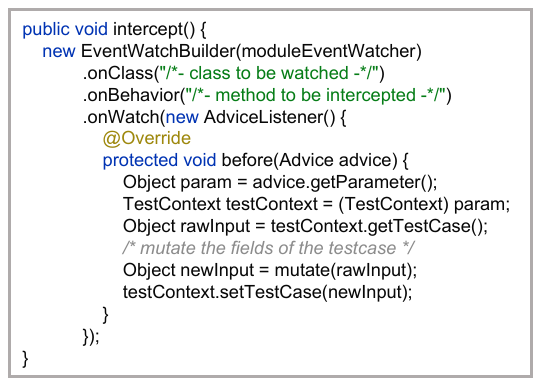}
    \caption{\textbf{A code example of sandbox interception.}}
    \Description{}
    \label{fig:sandbox-prototype}
\end{figure}

\subsection{Architecture of \textsc{SandBoxFuzz}} \label{sub:sandbox}
\textsc{SandBoxFuzz} is a solution upgraded from \textsc{FatFuzz} and fits well within the industrial context in Ant Group. Figure \ref{fig:SandboxFuzz} illustrates the architecture and workflow of \textsc{SandBoxFuzz}. 
To address the remaining challenges not addressed by \textsc{FatFuzz}, \textsc{SandBoxFuzz} features the following two further improvements:

\begin{itemize}
\item \textbf{Sandbox Interception.} Figure \ref{fig:sandbox-prototype} shows a simple code example of sandbox interception using JVM-Sandbox \cite{jvmsandbox}, a tool enabling aspect-oriented programming (AOP) for weaving new logics into specified methods. Target methods whose execution should be intercepted are referred as \textit{pointcuts}. To capture inputs at specific points within the lifecycle of test cases, \textsc{SandBoxFuzz} registers listeners that intercept fixed invocation points of \textit{iTest} as \textit{pointcuts}. 
When the specified methods of \textit{iTest} are invoked, the runtime arguments of test cases are intercepted and modified, before resuming the testing processes. As shown in \textit{Step 0} of Figure \ref{fig:SandboxFuzz}, similar to how \textsc{FatFuzz} executes multiple inputs within a single iteration, \textsc{SandBoxFuzz} ensures successful launches of the test execution processes by duplicating valid inputs from manual test suite. 
We refer to these inputs as dummy test placeholders. In \textit{Step 1}, these inputs are intercepted and processed at specific \textit{pointcuts}, with actual test inputs short-circuited. These inputs are produced by runtime mutator in \textit{Step 2} and have larger mutation space than file formats. By using dummy test placeholders, we can skip the operations for checking the validity of inputs and avoid input discard issues, given that the placeholder test inputs are \textit{syntactically valid}.
By default, \textsc{SandBoxFuzz} initializes 100 dummy test placeholders in the first iteration, and subsequently, the number of dummy test placeholders is prepared according to the number of required inputs. 
\item \textbf{Specification Mining.} \textsc{SandBoxFuzz} uses the input specifications to ensure that the dummy test placeholders are mutated into other valid inputs.
The specification provides information about the valid types and ranges of input values. To avoid the time-consuming manual process of specification configuration,
in \textsc{SandBoxFuzz}, we introduce specification mining, which is performed automatically at \textit{Step 1} on the intercepted input objects, as shown in Figure \ref{fig:SandboxFuzz}. Specifically, the structure of the input objects, their fields, and nested objects are extracted at runtime by Java's reflection APIs. 
This saves substantial manual work in configuring correct input specifications while setting up a runnable fuzzing pipeline.
\end{itemize}

As shown in Figure \ref{fig:SandboxFuzz}, the workflow of \textsc{SandBoxFuzz} includes the following stages:    

\begin{enumerate}
\item \textbf{Setup.} Set up the environments for \textsc{SandBoxFuzz}, apply instrumentation on compiled bytecode, switch to the target \textit{entry method} and prepare for dummy test placeholders, as shown in \textit{Step 0}. After these operations, the executable test suite is ready to be launched. 
\item \textbf{Specification Mining.} As shown in \textit{Round 0} at \textit{Step 1}, specification mining is applied on the inputs from manually written test cases upon sandbox interception.
The specifications are stored in a Redis database, and retrieved during input generation in the subsequent rounds.
\item \textbf{Fuzzing Iterations.} 
As shown in \textit{Step 2}, upon completing an iteration, the runtime mutator of \textsc{SandBoxFuzz} generates new inputs by mutating on existing seeds, stores them into a Redis database and updates the number of dummy test placeholders for the next iteration (return to \textit{Step 0}). Again at execution, as shown in \textit{Round 1 \& After} at \textit{Step 2}, these dummy inputs are replaced with new generated ones at specific \textit{pointcut}s. Outputs from \textit{iTest} are processed by a filter, which identifies any coverage updates or exceptions. An early stop strategy is performed where \textsc{SandBoxFuzz} stops the fuzzing pipeline once coverage is not updated for three iterations, since Ant Group limits the temporal and computational resources available for fuzzing.
\end{enumerate}

\begin{figure}[htbp]
    \centering
    \includegraphics[width=0.95\linewidth]{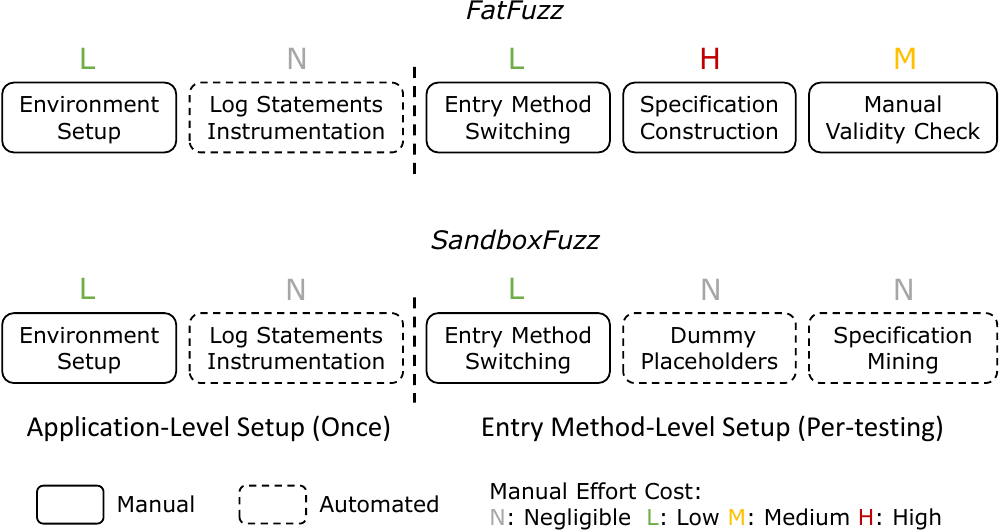}
    \caption{\textbf{Comparison of manual cost on preparation steps.}}
    \Description{}
    \label{fig:preparation}
\end{figure}

\subsection{\textsc{SandBoxFuzz} vs. \textsc{FatFuzz}} \label{sub:diff}

The dashed-line boxes in Figure \ref{fig:SandboxFuzz} show the differences of workflows between \textsc{SandBoxFuzz} (with \textit{blue} background) and \textsc{FatFuzz} (with \textit{green} background). As shown in \textit{Step a} and \textit{Step b}, \textsc{FatFuzz} relies on direct mutation on the configuration files. In contrast, using sandbox interception, \textsc{SandBoxFuzz} bypasses any input discard issues and replaces the dummy inputs at runtime. It produces a larger number of complex valid inputs in a larger mutation space by switching from raw files to memory objects and thus yields a better coverage than \textsc{FatFuzz}.

Figure \ref{fig:preparation} depicts in detail how \textsc{SandBoxFuzz} saves human efforts in applying the fuzzing pipeline. In practice, multiple configuration steps are required to get the fuzzing pipeline ready. All of them first go through an application-level setup process, including configuring the dependencies and environments for the fuzzer, and an instrumentation process with minimal human efforts. In \textit{entry method}-level setup, \textsc{FatFuzz} requires substantial human efforts in constructing an input specification.
On average, this time-consuming process takes a couple of hours for every single \textit{entry method}.
Instead, \textsc{SandBoxFuzz} avoids this overhead by mining specifications upon sandbox interception. 
By avoiding both potential and inevitable human errors, \textsc{SandBoxFuzz} typically reduces manual preparation time by hours, making it possible for a scalable deployment of the fuzzer within industrial environments.

\subsection{Input Generation} \label{sub:mutation}

As illustrated in \textit{Step 2} of Figure \ref{fig:SandboxFuzz}, at the end of each iteration, \textsc{SandBoxFuzz} collects the inputs that contribute to coverage as \textit{seed}s. Input generation is performed by applying mutation and crossover operators on them.

\begin{algorithm}[hbtp]
    \caption{Input Generation}
    \label{IG}
    \begin{algorithmic}[0]
    
    \renewcommand{\algorithmicrequire}{\textbf{Input:}}
    \renewcommand{\algorithmicensure}{\textbf{Output:}}
    
    \Require {\textit{evolution pool}, \textit{top-K pool} and \textit{maxCoverage}.}
    \Ensure {$S$, the fresh inputs for next iteration.}

    \State $S$ $\leftarrow$ $\{\}$
    
    \For{$seed \in $ \textit{evolution pool}}
        \State \textit{numOfChildren} $\leftarrow$ $\textit{baseNum} * \textit{coverage(seed)} / \textit{maxCoverage}$
        \If{$seed$ is favoured}
            \State \textit{numOfChildren} $\leftarrow$ $\textit{numOfChildren} * \textit{favourFactor}$
        \EndIf
        \For{$1 \leq i \leq \textit{numOfChildren}$}
            \State $S$ $\leftarrow$ $S \cup \textit{Mutation}\textit{(seed)}$ 
        \EndFor
    \EndFor

    \For{$\{seed1, seed2\} \in $ \textit{allPairsOf}\textit{(top-K pool)}}
        \State $S$ $\leftarrow$ $S \cup \textit{Crossover}\textit{(seed1, seed2)}$ 
    \EndFor
    
    \State \Return $S$
    
    \end{algorithmic}
\end{algorithm}

Algorithm \ref{IG} describes the input generation process when finishing one iteration. Two pools are maintained for holding valid seeds: \textit{evolution pool} and \textit{top-K pool}. At the beginning of fuzzing, both pools are empty since no inputs get executed and output. The \textit{evolution pool} stores inputs that contribute to coverage. 
Following the implementation of JQF \cite{JQF}, any inputs that double the coverage count on covered branches or cover uncovered branches are saved to \textit{evolution pool}. Note that the inputs that cover previously uncovered branches are favoured seeds. The \textit{top-K pool} always retains inputs with top K coverage and will replace less favored inputs.

\begin{algorithm}[hbtp]
    \caption{Crossover Operator}
    \label{CO}
    \begin{algorithmic}[0]
    
    \renewcommand{\algorithmicrequire}{\textbf{Input:}}
    \renewcommand{\algorithmicensure}{\textbf{Output:}}
    
    \Require {$O_{1}$ and $O_{2}$, the input objects.}
    \Ensure {$O^{*}$, the output object.}

    \State $O^{*}$ $\leftarrow$ \textit{newDefaultInstanceFrom}($O_{1}$)
    \State $F_{P}$, $F_{NP}$ $\leftarrow$ \textit{getPrimitiveFields}($O_{1}$), \textit{getNonPrimitiveFields}($O_{1}$)
    \State $F_{P1}, F_{P2}$ $\leftarrow$ select a middle point, parse $F_{P}$ into two sets
    \For{$field \in F_{P1}$}
        \State $O^{*}$.\textit{setValue}($field$, $O_{1}$.\textit{getValue}($field$)) 
    \EndFor
    \For{$field \in F_{P2}$}
        \State $O^{*}$.\textit{setValue}($field$, $O_{2}$.\textit{getValue}($field$)) 
    \EndFor
    \For{$field \in F_{NP}$}
        \State $O_{1}^{*}$, $O_{2}^{*}$ $\leftarrow$ $O_{1}$.\textit{getValue}($field$), $O_{2}$.\textit{getValue}($field$)
        \State $O^{*}$.\textit{setValue}($field$, \textit{Crossover}\textit{($O_{1}^{*}$, $O_{2}^{*}$)})
    \EndFor
    \State \Return $O^{*}$
    
    \end{algorithmic}
\end{algorithm}

Different mutation operators are applied on the seeds from these two pools. For mutation operators, new inputs are produced from seeds of \textit{evolution pool} through mutations, where a subset of fields is randomly selected and modified based on the input structures outlined in previously analyzed specifications.
Crossover operators are used to generate more diverse and effective test cases. As described in Algorithm \ref{CO}, multi-level crossover is applied between each pair of parent inputs from \textit{top-K pool} to generate new offspring inputs. When self-referencing objects or recursive structures are detected, the mutation is terminated.
Note that using input specifications while tackling non-primitive and private properties can prevent causing invalid input objects.

\section{EVALUATION} \label{evaluation}

In this section, we conduct extensive experiments to answer the following research questions:

\begin{itemize}
\item \textbf{RQ1:} How effective is \textsc{SandBoxFuzz} at fuzzing target \textit{entry method}s, as compared to \textsc{FatFuzz}?
\item \textbf{RQ2:} What is the performance and overhead of \textsc{SandBoxFuzz}, as compared to \textsc{FatFuzz}?
\item \textbf{RQ3:} Can \textsc{SandBoxFuzz} be efficiently deployed on new \textit{entry method}s by developers in an industrial environment?
\end{itemize}

To this end, we deployed \textsc{SandBoxFuzz} on Ant Group's real, large-scale microservices systems. In our evaluation, these target \textit{entry method}s come from 4 applications, the information of which is shown in Table \ref{tab:appinfo}. Columns named \textbf{\#LoC} and \textbf{Size} list the lines of code and number of bytes of repository size in each application. \textbf{\#NoM} lists the total number of methods and \textbf{\#NoB} lists the total number of basic blocks (also denotes the number of branches) in all CFGs of the application. \textbf{\#NoTC} lists the number of test cases in existing manual test suite and \textbf{\#NoEM} lists the number of unique \textit{entry method}s to which they belong to. 
To assess \textsc{SandBoxFuzz}'s effectiveness, we use both \textsc{FatFuzz} and the manually-written test cases for baseline comparisons.

\begin{table}[]
\centering
\setlength{\tabcolsep}{1.3mm}
\caption{\textbf{Application information.}}
\label{tab:appinfo}
\begin{threeparttable}
\begin{tabular}{|c|c|c|c|c|cc|}
\hline
\multirow{2}{*}{\textbf{App}} &
  \multirow{2}{*}{\textbf{\#LoC}} &
  \multirow{2}{*}{\textbf{Size}} &
  \multirow{2}{*}{\textbf{\#NoM}} &
  \multirow{2}{*}{\textbf{\#NoB}} &
  \multicolumn{2}{c|}{\textbf{MTS\tnote{*}}} \\ \cline{6-7} 
   &        &       &       &       & \multicolumn{1}{c|}{\textbf{\#NoTC}} & {\textbf{\#NoEM}} \\ \hline
A1 & 517.2K & 592.0MB & 28.3K & 57.1K & \multicolumn{1}{c|}{17}           & {6}            \\ \hline
A2 & 132.8K & 76.3MB  & 5.8K  & 10.1K & \multicolumn{1}{c|}{251}          & {59}           \\ \hline
A3 & 376.4K & 437.6MB & 38.5K & 71.8K & \multicolumn{1}{c|}{1428}         & {332}          \\ \hline
A4 & 212.6K & 119.4MB & 23.2K & 44.6K & \multicolumn{1}{c|}{518}          & {103}          \\ \hline
\end{tabular}
\begin{tablenotes}
    \item[*] Manual Test Suite
\end{tablenotes}
\end{threeparttable}
\end{table}

\subsection{Experiment Setup}
Evaluations are conducted on 4 elastic cloud instances of an internal cluster, each of which can serve one application and is equipped with 32 2.7GHz cores and 16GB of RAM. 
Execution process takes place in a staging environment where different versions of microservices may exhibit different behaviors due to deployments of software updates.
Experiments across different configurations and baselines on each \textit{entry method} were conducted concurrently in the same period to mitigate the effect of behavioral deviations.

Both \textsc{SandBoxFuzz} and \textsc{FatFuzz} use the same initial seeds from manual test suite to generate a fixed number (100) of inputs in the first iteration. Subsequent inputs are generated from seeds of two seed pools. \textit{baseNum} and \textit{favourFactor} used in Algorithm \ref{IG} are set to 10 and 5 respectively. The size of \textit{top-K} pool is set to 10.
The fuzzing loop is terminated if coverage status is not updated for 3 iterations. 
Following the focus of the quality-control team at Ant Group on 19 common critical runtime exceptions, we evaluated the fuzzers by identifying and tallying these target exceptions using the execution logs obtained from \textit{iTest}. 

\begin{table*}[!ht]
\centering
\setlength{\tabcolsep}{0.9mm}
\caption{\textbf{Experiment results for RQ1.}}
\label{tab:RQ1}
\begin{threeparttable}
\begin{tabular}{cccccccccccccccccccc}
\Xhline{1.3pt}
\multirow{2}{*}{\textbf{Entry Method}} & \multicolumn{6}{c}{\textbf{SandBoxFuzz}} && \multicolumn{6}{c}{\textbf{FatFuzz}} && \multicolumn{4}{c}{\textbf{MTS}\tnote{*}} \\ \cline{2-7} \cline{9-14} \cline{16-19} & \textbf{\#CM} & \textbf{\#CB} & \textbf{\#VUL} & \textbf{Cases} & \textbf{Seeds} & \textbf{Itrs} && \textbf{\#CM} & \textbf{\#CB} & \textbf{\#VUL} & \textbf{Cases} & \textbf{Seeds} & \textbf{Itrs} && \textbf{\#CM} & \textbf{\#CB} & \textbf{\#VUL} & \textbf{Cases} \\ \Xhline{1.3pt}
EM1 & 492 & 1013 & 193 & 5749 & 30 & 9 &  & 503 & 1018 & 0 & 3511 & 21 & 7 &  & 360 & 709 & 0 & 1 & \\ 
EM2 & 637 & 1207 & 792 & 8644 & 73 & 7 &  & 534 & 1017 & 0 & 4868 & 30 & 7 &  & 426 & 747 & 0 & 1 &\\ 
EM3 & 512 & 1008 & 91 & 4075 & 40 & 6 &  & 484 & 943 & 0 & 3842 & 26 & 7 &  & 383 & 693 & 0 & 1 & \\ 
EM4 & 473 & 956 & 35 & 7085 & 32 & 9 &  & 447 & 885 & 0 & 5278 & 23 & 9 &  & 402 & 760 & 0 & 9 &\\ 
EM5 & 752 & 1231 & 1 & 7058 & 97 & 8 &  & 740 & 1188 & 37 & 2000 & 39 & 5 &  & 713 & 1123 & 1 & 7 &\\ 
EM6 & 276 & 339 & 594 & 2209 & 10 & 8 &  & 195 & 227 & 0 & 592 & 3 & 4 &  & 81 & 112 & 0 & 1 &\\ 
EM7 & 1289 & 1766 & 181 & 2014 & 18 & 5 &  & 1295 & 1776 & 0 & 1906 & 15 & 6 &  & 1526 & 2033 & 0 & 4 & \\ 
\hline \textit{\textbf{total}} & \textbf{4431} & \textbf{7520} & \textbf{1887} & \textbf{36834} & \textbf{300} & - &  & 4198 & 7054 & 37 & 21997 & 157 & - &  & 3891 & 6177 & 1 & 24 & \\ 
\textit{\textbf{diff}} & \textbf{+13.9\%} & \textbf{+21.7\%} & \textbf{+1886} & - & - & - &  & +7.9\% & +14.2\% & +36 & - & - & - &  & - & - & - & - \\ 
\Xhline{1.3pt}
\end{tabular}
\begin{tablenotes}
    \item[*] Manual Test Suite
\end{tablenotes}
\end{threeparttable}
\end{table*}

\begin{table}[]
\centering
\setlength{\tabcolsep}{1mm}
\caption{\textbf{Types of exceptions present in results of RQ1.}}
\label{tab:exception}
\begin{tabular}{ll}
\Xhline{1.3pt}
\textbf{Exception Name} & \\ \Xhline{1.3pt}
NullPointerException & ArrayIndexOutOfBoundsException \\
NumberFormatException & IllegalStateException \\
IllegalArgumentException & SQLException\\
RuntimeException & IOException \\
ClassCastException & \\ \Xhline{1.3pt}
\end{tabular}
\end{table}

\begin{table}[htbp]
\centering
\setlength{\tabcolsep}{0.9mm}
\caption{\textbf{Results of performance evaluation for RQ2.}}
\label{tab:RQ2-1}
\begin{tabular}{ccccccccc}
\Xhline{1.3pt}
\multirow{2}{*}{\textbf{Entry Method}} & \multicolumn{3}{c}{\textbf{SandBoxFuzz}} && \multicolumn{3}{c}{\textbf{FatFuzz}} \\ \cline{2-4} \cline{6-8} 
 & \textbf{\#CM} & \textbf{\#CB} & \textbf{\#VUL} && \textbf{\#CM} & \textbf{\#CB} & \textbf{\#VUL} & \\ \Xhline{1.3pt}
EM1 & 424 & 911 & 1 &  & 414 & 875 & 0 &  \\ 
EM2 & 549 & 1062 & 203 &  & 463 & 900 & 0 &  \\ 
EM3 & 435 & 849 & 0 &  & 415 & 823 & 0 &  \\ 
EM4 & 427 & 870 & 28 &  & 385 & 789 & 0 &  \\ 
EM5 & 749 & 1215 & 4 &  & 740 & 1188 & 47 &  \\ 
EM6 & 276 & 339 & 675 &  & 195 & 227 & 0 &  \\ 
EM7 & 1289 & 1768 & 287 &  & 1289 & 1778 & 0 &  \\ 
\hline \textit{\textbf{total}} & \textbf{4149} & \textbf{7014} & \textbf{1198} &  & 3901 & 6580 & 47 &  \\ 
\Xhline{1.3pt}
\end{tabular}
\end{table}

\subsection{RQ1: The Effectiveness of \textsc{SandBoxFuzz}}

To evaluate the effectiveness of \textsc{SandBoxFuzz},
we have conducted fuzzing experiments both for \textsc{SandBoxFuzz} and \textsc{FatFuzz} on seven selected \textit{entry method}s, with non-trivial execution logics and perceived to be error-prone by their developers.

Table \ref{tab:RQ1} presents the results of our experiments. 
The column \textbf{\#CM} represents the covered unique methods.
\textbf{\#CB} represents the covered unique basic blocks, which also indicates the branch coverage. 
\textbf{\#VUL} lists the number of test cases that have discovered targeted exceptions from the log results of \textit{iTest}. 
\textbf{Cases} lists the total number of test cases being executed and \textbf{Seeds} lists the number of test cases that contribute to coverage. 
\textbf{Itrs} shows how many iterations are executed.
\textbf{MTS} lists the execution results of manual test suite, which are crucial cases that developers concerned and designed based on historical traces.

Compared with \textsc{FatFuzz}, \textsc{SandBoxFuzz} improves 6.0\% and 7.5\% in method and branch coverage respectively, as well as yields 1,850 more test cases in which the target exceptions were raised.
For both approaches, 9 out of 19 targeted exceptions in total were successfully triggered by the fuzzers, as detected in the execution logs from \textit{iTest}. 
Table \ref{tab:exception} lists the types of these exceptions. Among them, \textit{NullPointerException} accounts for the majority of the results with three quarters of the occurrences. The first four types of exceptions in the first column of Table \ref{tab:exception} account for over 95\%. 
Note that our results do not include exceptions caused by \textit{syntactically invalid} inputs, which are detected before actual execution of test cases by checking against custom exceptions within \textit{iTest}.

The results suggest that \textsc{SandBoxFuzz} is more effective than \textsc{FatFuzz} in improving coverage and discovering potential vulnerabilities. This is because \textsc{SandBoxFuzz} mitigates the constraints imposed by \textit{iTest} that \textsc{FatFuzz} fails to address, and is able to perform more flexible mutations. Thus, \textsc{SandBoxFuzz} has a higher throughput, producing more inputs with higher quality, 
contributing in greater coverage and yielding more exceptions. The only exception in EM5 was that 37 input validation errors were observed. \textsc{FatFuzz}, which relies on FastJson \cite{fastjson} for Java object transformation, caused issues, while \textsc{SandBoxFuzz} dynamically mutates input parameters on the fly to avoid such problems.

\textsc{SandBoxFuzz} has also demonstrated more effective fuzzing, compared with the manual test suite. 
In our analysis, we find that developers
usually write limited inputs only to check whether the code is functionally correct. 
They do not write test cases that expose potential vulnerabilities for a variety
of reasons, including the amount of time required and the error-prone  process for preparing valid test cases that test deeper logics. 
For \textsc{SandBoxFuzz}, specifications are mined, and the fuzzer can run without manual preparation.

\subsection{RQ2: The Performance and Overhead of \textsc{SandBoxFuzz}}

\begin{figure}[]
    \centering
    \includegraphics[width=1\linewidth]{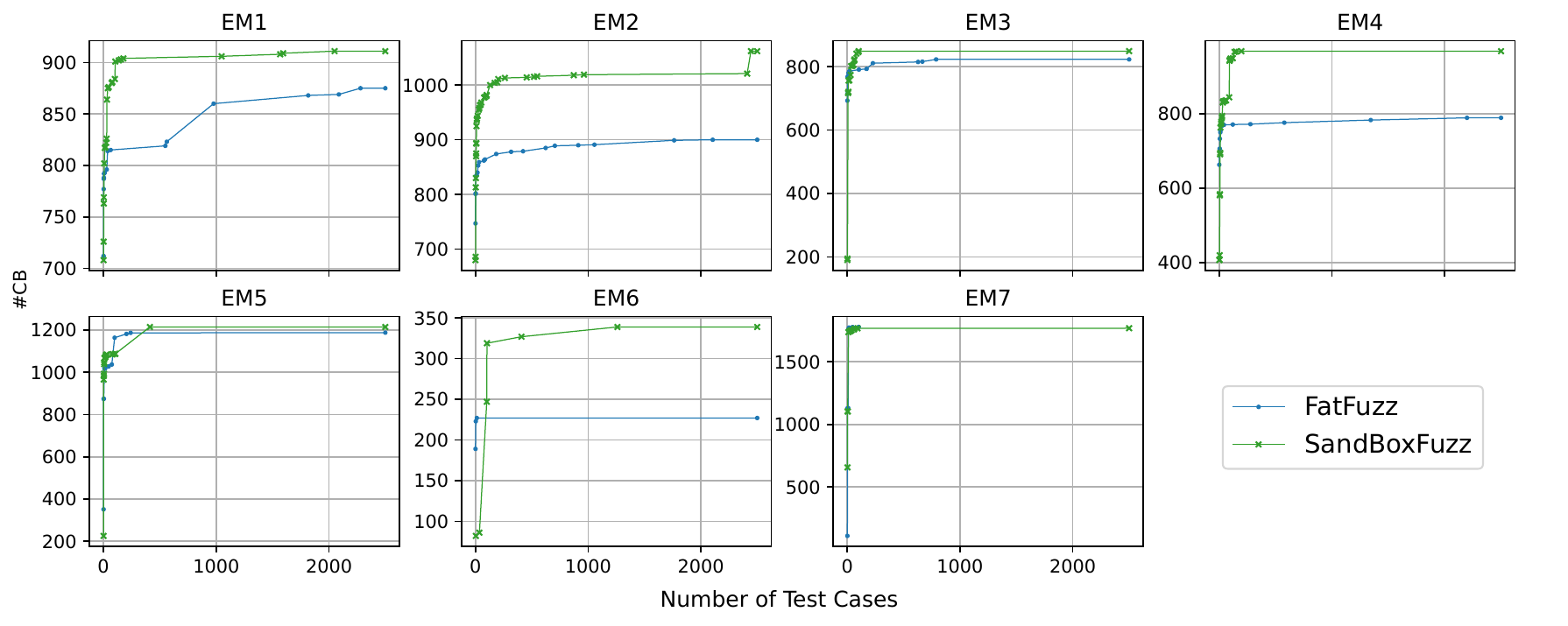}
    \caption{\textbf{Number of covered blocks versus executed test cases for RQ2.}}
    \Description{}
    \label{fig:RQ2overtime}
\end{figure}

To understand the performance and overhead of \textsc{SandBoxFuzz}, we conduct a comparison study with additional settings on the same \textit{entry method}s listed in RQ1.
Each iteration strictly generates 400 test cases and the fuzzing process is stopped once a total of 2,500 test cases has been accumulated.
Table \ref{tab:RQ2-1} lists the final results and Figure \ref{fig:RQ2overtime} shows how the number of covered basic blocks change with the number of executed test cases.
The results
show that \textsc{SandBoxFuzz} still yields better coverage and more exceptions than \textsc{FatFuzz}. 

\begin{figure}[htbp]
    \centering
    \Description{}
    \begin{subfigure}[t]{0.4\textwidth}
        \centering
        \includegraphics[width=\textwidth]{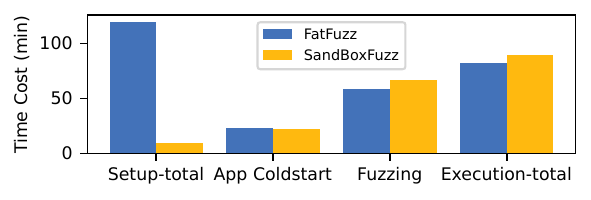}
        \caption{\textbf{Average time cost comparison across different phases.}}
        \label{fig:RQ2-Big-1}
    \end{subfigure}
    \vfill
    \begin{subfigure}[t]{0.4\textwidth}
        \centering
        \includegraphics[width=\textwidth]{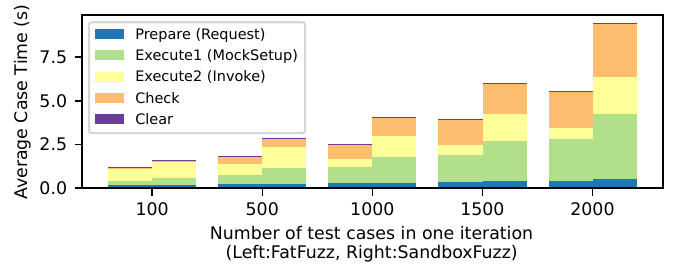}
        \caption{\textbf{Breakdown of execution results in \textit{iTest}.}}
        \label{fig:RQ2-Big-2}
    \end{subfigure}
    \caption{\textbf{Measurement results of time cost and overhead.}}
    \label{fig:RQ2-Big}
\end{figure}

Figure \ref{fig:RQ2-Big-1} shows the results of time cost in these \textit{entry method}s. Both approaches have the same time cost in cold start processes and have a comparable execution time. Most importantly, \textsc{SandBoxFuzz} has a much lower manual preparation cost (10 minutes vs. 2 hours), and consequently it reduces overall time cost by 50\%.
To understand the additional overhead introduced by \textsc{SandBoxFuzz}, we studied the worst case by performing execution breakdown on target \textit{entry method} (EM5), which has the biggest deviation of execution cost between two approaches. As shown in Figure \ref{fig:RQ2-Big-2}, \textsc{SandBoxFuzz} brings additional overhead at different points of \textit{iTest}, each with hundreds of milliseconds, due to setting up the sandbox agent. The mock setup, invoke, and check processes involve traversal of input objects, so \textsc{SandBoxFuzz} spends more time since it produces more complex inputs. There is also a linear increase in costs associated with request preparation, mock setup and validation processes, as they involve file operations with cost directly related to the size of files.
By including a proper number of inputs in each iteration, the additional overhead introduced by \textsc{SandBoxFuzz} is acceptable in terms of overall fuzzing time shown in Figure \ref{fig:RQ2-Big-1}, where 400 test cases are included in each iteration.

\begin{figure*}[htbp]
    \centering
    \includegraphics[width=1\linewidth]{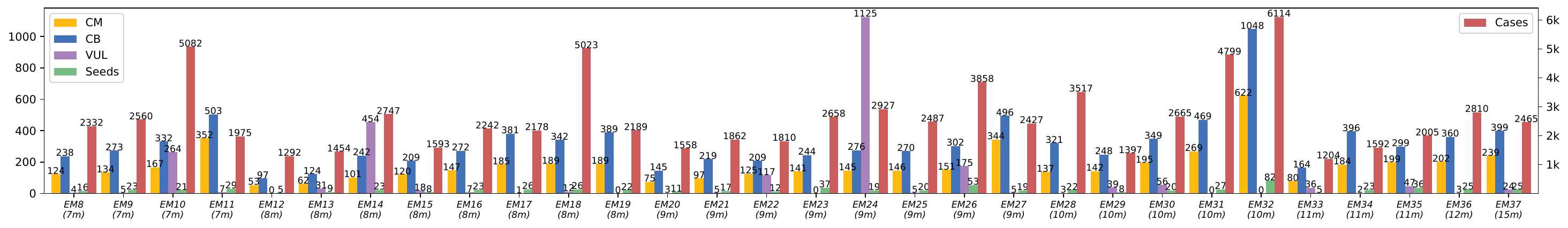}
    \caption{\textbf{Experiments results for RQ3.}}
    \Description{}
    \label{fig:RQ3-2&3}
\end{figure*}

\subsection{RQ3: The Scalability of \textsc{SandBoxFuzz}}

To answer RQ3, we study how easily \textsc{SandBoxFuzz} can be scaled and applied to other systems. Specifically, we invited two developers unfamiliar with \textsc{SandBoxFuzz}, to deploy \textsc{SandBoxFuzz} on 30 new \textit{entry method}s from applications listed in Table \ref{tab:appinfo} and performed assessment on the manual cost and fuzzing effectiveness. 

Our user study shows that setting up each of the 30 new \textit{entry method}s manually takes between 7 and 15 minutes, with an average of 9 minutes. These times are indicated on the x-axis labels of
Figure \ref{fig:RQ3-2&3}.
Locating and formatting existing valid inputs of target \textit{entry method}, along with debugging errors, account for the majority of the manual preparation time cost.
Moreover, the time expense can be further reduced if developers gain familiarity with \textsc{SandBoxFuzz}. The processes of learning \textsc{SandBoxFuzz} and setting up the corresponding environments are included for time budget as well, but are incurred only once per application. 
In comparison, the preparation process of the baseline \textsc{FatFuzz} is error-prone, requiring trial-and-error. Manually constructing correct input specifications is difficult, with an average cost of several hours. In their initial use of \textsc{FatFuzz}, the developers tried to set it up for several new \textit{entry method}s but were unable to complete them due to the complex preparation steps. The bars of Figure \ref{fig:RQ3-2&3} present the fuzzing results. In total, 2,448 out of 78,822 cases have uncovered target exceptions, demonstrating the effectiveness. For these 30 \textit{entry method}s, on average, the processes of application cold start and execution cost 130 minutes in total, with only 9 minutes required for the manual preparation process.

\section{DISCUSSION} \label{discussion}

In this section, we discuss the lessons learned and threats to validity.

\subsection{Lessons Learned}
From our real practice, we learned the following lessons.

\textbf{It is challenging but necessary to build fuzzers that do not depend on a tight feedback loop and full permissions.}
Grey-box fuzzers are usually much more effective than black-box fuzzers, but require feedback from efficient executions. 
In large-scale industrial systems, executions may not be fast due to invocations to numerous microservices components and the multiple costly reboot processes, while the test environments usually provide limited or even no valuable feedback.
On the other hand, black-box fuzzers without any guidance underperform grey-box fuzzers.
There's a need for research to investigate methods of guiding the fuzzing loop effectively with minimal feedback, especially under conditions where access and modifications to the test environments are prohibited due to confidentiality considerations.

\textbf{Instrumentation may not be cheap.}
Fuzzers require lightweight instrumentation to obtain feedback. 
A standard instrumentation process involves setting up a Java agent at runtime to intercept execution and collect coverage metrics. 
However, in industrial testing, the coverage states, even in a Java agent, may be lost during the restarts of testing pipeline. 
Another challenge is that many execution paths are less interesting to test within large-scale systems, but 
the coverage of these paths, related to setting up and loading services, still contribute to the coverage results collected.
There is value for future research on a lightweight instrumentation that still provides accurate coverage feedback.

\textbf{Specifications mining can bootstrap the fuzzer.}
Input specifications allow the creation of \textit{syntactically valid} inputs, and can be inferred from either user knowledge or instances of input objects at runtime. 
Note that writing more effective custom generators (as in JQF\cite{JQF} and Zest\cite{zest}) for valid input objects according to input specifications would require much more manual effort.
In addition, for industrial microservices software, logics are embedded deeply in program with complex dependencies. 
Future research should investigate if custom input generators and specifications can help in testing deeper parts of the components. 
Tools such as LLM can facilitate the construction of more powerful specifications, generating more effective inputs than rule-based ones.

\textbf{The manual effort required to set up a fuzzer has to be mitigated.}
Despite the benefits of fuzzing, our experiences show that developers are reluctant to adopt fuzzing if they find a long preparation time. The preparation process includes multiple steps that requires manual work such as pipeline setup, specification construction and additional validity checks. 
Given the rapid convergence of coverage (thus short fuzzing footprints) in FinTech systems, prolonged manual setup processes present significant challenges in achieving scalable fuzzing.
Meanwhile, users of fuzzers may lack knowledge of target program and test environments, adding difficulty in correctly setting up the fuzzer. It's recommended to automate as many of the preparation steps as possible, and offer a solution with one-click deployment for users of fuzzers.

\subsection{Threats to Validity}
Several potential threats to validity exist in our approach due to the complexity and dynamic nature of microservices architectures. One significant threat to validity is the temporal variability of microservices. System behavior changes after software updates or deployments, leading to fluctuations in test coverage and fuzzing results. While previous work \cite{di2024microfuzz} mitigated this by using mocking to freeze prior versions of microservices, our setting—despite being within the same Ant Group ecosystem—utilizes different test frameworks such as \textit{iTest}, which does not support version freezing. 
Instead, we mitigated this issue through standardized service update schedules during our experiments.
Future work could dynamically freeze stable versions of microservices using aspect-oriented programming techniques, providing better control over temporal variability. 

Another concern involves instrumentation failures, particularly inconsistencies in bytecode transformations (e.g., .\texttt{class} to .\texttt{jimple} and back). These issues, exacerbated by frameworks like Spring that rely on annotations such as \texttt{@Autowired}, can cause crashes when the application runs. 
To mitigate this, we excluded problematic modules from instrumentation and refined the transformation process to reduce errors.
Using a single baseline is a concern, but our pilot study shows that \textsc{FatFuzz}, built on top of the SOTA fuzzing, is best suited for the Ant Group environment and specifically designed for it.
External validity is another concern, as the generalizability of our approach may be limited by the specific microservices architectures and configurations. 
To enhance external validity, we selected diverse entry methods to test representative of industry practices as covered in RQ3.

\section{RELATED WORK} \label{relatedwork}
Fuzzing has been widely applied in industrial settings to address challenges such as proprietary systems and large-scale distributed architectures. Researchers at Huawei \cite{liang2018fuzz} proposed a fuzzing framework combining static analysis with runtime instrumentation to enhance test generation efficiency. Their approach integrates runtime feedback to refine test cases dynamically and focuses on automated seed generation. In contrast, our work addresses corporate constraints in large-scale distributed systems, specifically resolving issues like error-prone and time-consuming preparation process of initial test cases. Our approach bypasses unnecessary validation and restrictions imposed by corporate test frameworks, allowing for effortless test case initialization. 

In Ant Group, MicroFuzz \cite{di2024microfuzz} leverages parallelism to accelerate fuzzing across multiple microservices, which complements our approach. However, our method directly addresses fuzzing slowness in the corporate settings without requiring parallelism. Similarly, FinHunter \cite{ding2024finhunter} generates valid test cases to bypass constraints of internal test frameworks, while our sandbox approach focuses on simplifying the process by intercepting real calling points for \textit{entry method}s and conducting specification mining and dynamic mutation at runtime. This makes the initial test case preparation almost effortless, requiring only a single existing manually written test case for specification mining in the iteration 0.

Zhang et al. \cite{zhang2023EvoMaster} enhanced EvoMaster, a white-box fuzzer, to evaluate RPC-based APIs using source code access, focusing on mocking external services to target core functionalities. Our grey-box fuzzing approach, in contrast, relies on runtime feedback for effective mutation and test coverage without requiring source code access. Zest \cite{zest}, a structured input fuzzer, inspired the development of \textsc{FatFuzz} at Ant Group, targeting microservices architectures. While \textsc{FatFuzz} mitigates test framework overhead through naive iterative test case generation, it does not fully resolve the issues of manual test case preparation and framework-imposed slowness. \textsc{SandBoxFuzz} fundamentally addresses these challenges by expanding the mutation space using JVM-Sandbox, eliminating the need for manual initialization, and improving fuzzing efficiency in industrial scenarios.

\section{CONCLUSION} \label{conclusion}
We propose a scalable and efficient grey-box fuzzing technique, \textsc{SandBoxFuzz}, designed for ultra-large microservices systems, addressing critical constraints, such as limited permissions, restrictive test environments, and input construction challenges. Leveraging aspect-oriented programming, \textsc{SandBoxFuzz} bypasses organizational limitations while maintaining compliance with data security regulations. Key insights include the amortization of startup costs, bytecode instrumentation, and the use of  runtime specification mining to significantly reduce manual effort. 
\textsc{SandBoxFuzz} improves branch coverage and exceptions caught, reducing setup time from hours to minutes, demonstrating its industrial scalability. It also provides detailed coverage information, enabling coverage-guided fault localization in large-scale, constrained industry systems.

\begin{acks}
This work is partially supported by the Tongji University Medicine-X Interdisciplinary Research Initiative under grant 2025-0553-YB-08, the National Natural Science Foundation of China under Grant 62372330, the Australian Research Council (ARC) through a Future Fellowship (FT240100269), a Discovery Project (DP210102447) and a Linkage Project (LP190100676).
\end{acks}

\clearpage
\balance
\bibliographystyle{ACM-Reference-Format}










\end{document}